\documentclass{article}
\usepackage{graphicx}
\usepackage{amsmath}
\usepackage{amsfonts}
\usepackage{amssymb}
\begin{document}

\title{Low-energy M\"{o}ller scattering in a Maxwell-Chern-Simons Lorentz-violating
planar model}
\author{M. M. Ferreira Junior\thanks{e-mail: manojr@ufma.br}\\\textit{Universidade Federal do Maranh\~{a}o (UFMA)}, \\Departamento de F\'{i}sica, Campus Universit\'{a}rio do Bacanga, S\~{a}o\\Luiz - MA, 65085-580 - Brazil.}
\maketitle
\begin{abstract}
One starts from a planar Maxwell-Chern-Simons model endowed with a
Lorentz-violating term. The Dirac sector is introduced exhibiting a Yukawa and
a minimal coupling with the scalar scalar and the gauge fields, respectively.
One then evaluates the electron-electron interaction as the Fourier transform
of the M\"{o}ller scattering amplitude carried out in the non-relativistic
limit. In the case of a purely time-like background, the interaction potential
can be exactly solved, exhibiting a typical massless behavior far from the
origin. The scalar interaction potential is always attractive whereas the
gauge intermediation may also present attraction even when considered in the
presence of the centrifugal barrier and the $A^{2}$ term. Such a result is a
strong indication that electron-electron bound states\ may appear in this
theoretical framework.
\end{abstract}

\section{Introduction}

In the beginning 90%
%TCIMACRO{\UNICODE{0xb4}}%
%BeginExpansion
\'{}%
%EndExpansion
s, the M\"{o}ller scattering was adopted as a theoretical tool to investigate
the possible formation of electron-electron bound states in the context of a
Maxwell-Chern-Simons electrodynamics \cite{MCS}. According to this procedure,
one starts from the scattering amplitude (carried out at tree-level) to obtain
the electron-electron interaction potential (Born approximation). As a settled
down result, it was observed that the potential may come out negative whenever
the topological mass exceeds the electron mass $\left(  s>m_{e}\right)  $,
condition which is particularly discouraging in relation to the possibility of
applying this kind of model to some condensed matter systems, where one
usually deals with low-energy excitations. The introduction of the Higgs
sector, arising from the spontaneous symmetry breaking \cite{Tese}, has shown
to be a theoretical factor able to provide a scalar attractive interaction.
The overall potential, consisting in the sum of the gauge and scalar
contributions, may then be negative independently of the condition $s>m_{e}$,
a necessary premise for the formation of Cooper pairs in the context of
low-energy systems.

In the latest years, Lorentz-violating theories have been in focus of
intensive investigation \cite{Kostelec},\cite{Colladay}. In a recent work, a
planar Lorentz-violating electrodynamics \cite{Manojr1} was derived from the
dimensional reduction of a Maxwell electrodynamics supplemented with the
Carroll-Field-Jackiw (CFJ) term \cite{Jackiw}. The consistency of this model
has already been analyzed, revealing a model globally stable, causal and
unitary for both time- and spacelike backgrounds \cite{Manojr1}. The fact that
the unitarity is assured makes feasible, at principle, the consistent
quantization of this model, which sets it up as a candidate to be applied to
situations where the quantization of the modes is a real condition (such as
some condensed matter phenomena). In a posterior investigation \cite{Manojr2},
the equations of motion (for the field strengths and potentials) corresponding
to this planar model were determined and solved in the static regime. The
results obtained differ from the solutions of a pure MCS electrodynamics by
background-depending corrections, which amount to relevant qualitative
modifications. Indeed, the solutions have exhibited a typical massless
behavior (in the electric sector) for the case of a timelike background and
anisotropic behavior for the case of a spacelike background. It was also
reported the possibility of obtaining an attractive electron-electron
interaction as a consequence of the existence of\ well region in the behavior
of the scalar potential $\left(  A_{0}\right)  $.

Lorentz covariance is certainly an essential feature of any relativistic
system, mainly for ensuring the equivalence between all inertial frames. Once
Lorentz symmetry is broken, such equivalence is lost, and each inertial frame
starts to notice a different physics. It is a well known fact that condensed
matter systems (CMS) are not endowed with Lorentz covariance, but with Galileo
one, which holds as a genuine symmetry in the domain of isotropic low-energy
systems. Having in mind that a CMS may be addressed as the low-energy limit of
a relativistic model, there follows a straightforward correspondence between
the breakdown of Lorentz and Galileo symmetries, in the sense that a CMS
with\ violation of Galileo symmetry may have as counterpart a relativistic
system endowed with breaking of Lorentz covariance.\thinspace\ 

Theoretical planar models able to provide attractive $e^{-}e^{-}$
interaction\ potentials are relevant in the sense they may constitute a
suitable framework to address the condensation of Cooper pairs, a fundamental
characteristic of superconducting systems. Another well defined feature of a
planar high-Tc superconductor concerns the symmetry of the order parameter
(standing for the Cooper pair), which is described in terms of a spatially
anisotropic d-wave \cite{OP}. A theoretical framework able to provide an
anisotropic $e^{-}e^{-}$ interaction is the first step to the achievement of
anisotropy for the order parameter. This is exactly the expected result to be
obtained in the case a pure spacelike background, where the $e^{-}e^{-}$
scattering potential may be identified with the one evaluated in the context
of a CMS endowed with a privileged direction in space. Therefore, once an
anisotropic CMS constitutes\ an example where the breakdown of the Galileo
symmetry takes place, such a system may be properly approached as the
low-energy limit of a Lorentz-violating electrodynamics in the presence of a
pure spacelike background.

Having as main motivation the results achieved in ref. \cite{Manojr2}, which
show that a fixed background induces sensitive effects at classical solutions,
in this work one investigates the tree-level behavior of two interacting
fermions in the context of a Lorentz-violating electrodynamics. By determining
of the $e^{-}e^{-}$ interaction potential, one can verify to what extent the
properties reported in the classical static analysis\ \cite{Manojr2} are
preserved in the context a dynamic\ evaluation. One can also study the
possibility of achieving an $e^{-}e^{-}$ interaction endowed with two relevant
features: attractiveness and anisotropy, relevant properties in
superconducting systems. Hence, the purpose is to carry out the $e^{-}e^{-}%
$\ interaction potential, exhibiting and stressing the corrections induced by
the fixed background on the pure Maxwell-Chern-Simons result. For that, one
first introduces the Dirac sector to the planar Lorentz-violating gauge model
derived in ref. \cite{Manojr1}. Taking into account the guidelines set up in
refs.\cite{MCS}, \cite{Tese}, one then proceeds to evaluate the M\"{o}ller
scattering amplitude from which one derives the $e^{-}e^{-}$ interaction
(according to the Born approximation). The potential here attained is composed
by two contributions, a scalar and a gauge one, since the $e^{-}e^{-}$
interaction is mediated by the massless scalar and the massive gauge fields.
The scalar potential, absent in the context of a pure MCS model, is always
negative, and may lead to a global attractive interaction regardless the sign
of the gauge contribution. In the case of the gauge potential, it presents
background-depending terms that imply qualitative modifications, such as the
possibility of being attractive for some parameters range, even when
considered in the presence of the centrifugal barrier and the low-energy
$A^{2}-$ Pauli term. Both the scalar and gauge potential possess a logarithm
dependence, which is compatible with a massless behavior far from the origin.

This paper is outlined as follows. In Sec. II, one briefly exhibits the
reduced model derived in ref. \cite{Manojr1}, supplemented by the fermion
field. In Sec. III, one presents the spinors which fulfill the two-dimensional
Dirac equation and are used to evaluate the scattering amplitude associated
with the Yukawa and the minimal interactions. In Sec. IV, the interaction
potential stemming form the scalar and gauge sectors are carried out, and the
results are discussed. In Sec.V, one presents the concluding remarks.

\section{Planar Lorentz-violating model}

The starting point is the planar Lagrangian\footnote{We adopt a
1+2-dimensional metric for space-time: $\eta_{\mu\nu}=(+,-,-).$} obtained from
the dimensional reduction of the CFJ-Maxwell electrodynamics \cite{Manojr1},
which consists in a Maxwell-Chern-Simons electrodynamics coupled to a massless
scalar field $\left(  \varphi\right)  $ and to a fixed background $\left(
v^{\mu}\right)  $ through a Lorentz-violating term. One then considers the
additional presence of a fermion field $\left(  \psi\right)  $ minimally
coupled to the gauge field $\left(  A_{\mu}\right)  $ at the same time that
exhibits a typical Yukawa coupling to the scalar field $\left(  \varphi
\right)  $:\ %

\begin{align}
\mathcal{L}_{1+2}  &  =-\frac{1}{4}F_{\mu\nu}F^{\mu\nu}+\frac{s}{2}%
\epsilon_{\mu\nu k}A^{\mu}\partial^{\nu}A^{k}-\frac{1}{2}\partial_{\mu}%
\varphi\partial^{\mu}\varphi+\varphi\epsilon_{\mu\nu k}v^{\mu}\partial^{\nu
}A^{k}-\frac{1}{2\alpha}\left(  \partial_{\mu}A^{\mu}\right)  ^{2}\nonumber\\
&  +{\overline{\psi}(i}{\rlap{\hbox{$\mskip4.5 mu /$}}D}-m_{e}){\psi}%
-y\varphi(\overline{\psi}\psi). \label{L1}%
\end{align}
Here, the covariant derivative, ${\rlap{\hbox{$\mskip4.5
mu /$}}D}\psi\equiv(\rlap{\hbox{$\mskip1
mu /$}}\partial+ie_{3}\rlap{\hbox{$\mskip3 mu /$}}A)\psi,$ states the minimal
coupling, whereas the term $\varphi(\overline{\psi}\psi)$ reflects the Yukawa
coupling. In ref. \cite{Manojr1}, the propagators of the scalar $\left(
\varphi\right)  $ and gauge $\left(  A_{\mu}\right)  $ fields were properly
evaluated as it appears below:
\begin{align}
\text{ }\langle A^{\mu}\left(  k\right)  A^{\nu}\left(  k\right)
\rangle\text{ }  &  =i\biggl\{-\frac{1}{k^{2}-s^{2}}\theta^{\mu\nu}%
-\frac{\alpha(k^{2}-s^{2})\boxtimes(k)+s^{2}\left(  v.k\right)  ^{2}}%
{k^{2}(k^{2}-s^{2})\boxtimes(k)}\omega^{\mu\nu}-\frac{s}{k^{2}(k^{2}-s^{2}%
)}S^{\mu\nu}\nonumber\\
&  +\frac{s^{2}}{(k^{2}-s^{2})\boxtimes(k)}\Lambda^{\mu\nu}-\frac{1}%
{(k^{2}-s^{2})\boxtimes(k)}T^{\mu}T^{\nu}+\frac{s}{(k^{2}-s^{2})\boxtimes
(k)}[Q^{\mu\nu}-Q^{\nu\mu}]\nonumber\\
&  +\frac{is^{2}\left(  v.k\right)  }{k^{2}(k^{2}-s^{2})\boxtimes(k)}%
[\Sigma^{\mu\nu}+\Sigma^{\nu\mu}]-\frac{is\left(  v.k\right)  }{k^{2}%
(k^{2}-s^{2})\boxtimes(k)}[\Phi^{\mu\nu}-\Phi^{\nu\mu}]\biggr\},
\label{Prop_A}%
\end{align}%
\begin{equation}
\text{ }\langle\varphi\varphi\rangle\text{ }=\frac{i}{\boxtimes(k)}\left[
k^{2}-s^{2}\right]  , \label{Prop_phi}%
\end{equation}
where: $\boxtimes(k)=\left[  k^{4}-\left(  s^{2}-v.v\right)  k^{2}-\left(
v.k\right)  ^{2}\right]  $, and the 2-rank tensors are defined as follows:
\begin{align}
\theta_{\mu\nu}  &  =\eta_{\mu\nu}-\omega_{\mu\nu},\text{\ }\omega_{\mu\nu
}=\partial_{\mu}\partial_{\nu}/\square,\text{ }S_{\mu\nu}=\varepsilon
_{\mu\kappa\nu}\partial^{\kappa},\text{ }Q_{\mu\nu}=v_{\mu}T_{\nu},\text{ }\\
\text{\ }T_{\nu}  &  =S_{\mu\nu}v^{\mu},\text{ }\Lambda_{\mu\nu}=v_{\mu}%
v_{\nu},\text{ \ }\Sigma_{\mu\nu}=v_{\mu}\partial_{\nu},\text{ }\Phi_{\mu\nu
}=T_{\mu}\partial_{\nu}.
\end{align}

\section{The M\"{o}ller Scattering amplitude}

The two-particle interaction potential is given by the Fourier transform of
the two-particle scattering amplitude in the low-energy limit (Born
approximation). In the case of the nonrelativistic M\"{o}ller scattering, one
should consider only the t-channel (direct scattering) \cite{Sakurai} even for
indistinguishable electrons, since in this limit they recover the classical
notion of trajectory. From eq. (\ref{L1}), there follow the Feynman rules for
the interaction vertices: $V_{\psi\varphi\psi}=iy;V_{\psi A\psi}=ie_{3}%
\gamma^{\mu}$, so that the $e^{-}e^{-}$ scattering amplitude are written as:
\begin{align}
-i\mathcal{M}_{\varphi}  &  =\overline{u}(p_{1}^{^{\prime}})(iy)u(p_{1}%
)\left[  \langle\varphi\varphi\rangle\right]  \overline{u}(p_{2}^{^{\prime}%
})(iy)u(p_{2}),\label{A1}\\
-i\mathcal{M}_{A}  &  =\overline{u}(p_{1}^{^{\prime}})(ie_{3}\gamma^{\mu
})u(p_{1})\left[  \langle A_{\mu}A_{\nu}\rangle\right]  \overline{u}%
(p_{2}^{^{\prime}})(ie_{3}\gamma^{\nu})u(p_{2}), \label{A2}%
\end{align}
with $\langle\varphi\varphi\rangle$ and $\langle A_{\mu}A_{\nu}\rangle$ being
the scalar and photon propagators. Expressions (\ref{A1}) and (\ref{A2})
represent the scattering amplitudes for electrons of equal polarization
mediated by the scalar and gauge particles, respectively. The spinors $u(p)$
stand for the positive-energy solution of the Dirac equation $\left(
\rlap{\hbox{$\mskip1 mu /$}}p-m\right)  u(p)=0$. The $\gamma-$ matrices
satisfy the $so(1,2)$ algebra, $\left[  \gamma^{\mu},\gamma^{\nu}\right]
=2i\epsilon^{\mu\nu\alpha}\gamma_{\alpha}$, and correspond to the
(1+2)-dimensional representation of the Dirac matrices, that is, the Pauli
ones: $\gamma^{\mu}=(\sigma_{z},-i\sigma_{x},i\sigma_{y}).$ Regarding these
definitions, one obtains the spinors,%

\begin{equation}
u(p)=\frac{1}{\sqrt{N}}\left[
\begin{array}
[c]{c}%
E+m\\
-ip_{x}-p_{y}%
\end{array}
\right]  ,\text{ \ \ }\overline{u}(p)=\frac{1}{\sqrt{N}}\left[
\begin{array}
[c]{cc}%
E+m & -ip_{x}+p_{y}%
\end{array}
\right]  , \label{spinor}%
\end{equation}
which fulfill the normalization condition $\overline{u}_{+}(p)u_{+}(p)=1$
whenever the constant $N=2m(E+m)$ is adopted. The M\"{o}ller scattering should
be easily analyzed in the center of mass frame, where the momenta of the
incoming and outgoing electrons are read at the form: \ $P_{1}^{\mu
}=(E,p,0),P_{2}^{\mu}=(E,-p,0),P_{1}^{^{\prime}\mu}=(E,p\cos\theta,p\sin
\theta),P_{2}^{^{\prime}\mu}=(E,-p\cos\theta,-p\sin\theta)$\footnote{Using
this prescription and the 3-current definition, $j^{\mu}(p)=\overline
{u}(p^{^{\prime}})\gamma^{\mu}u(p),$ the current components can be then
explicitly written as: $j^{\left(  0\right)  }(p_{1})=j^{\left(  0\right)
}(p_{2})=\frac{1}{2m(E+m)}[(E+m)^{2}+p^{2}e^{-i\theta}],j^{\left(  1\right)
}(p_{1})=-j^{\left(  1\right)  }(p_{2})=\frac{p}{2m}(1+e^{i\theta}),j^{\left(
2\right)  }(p_{1})=-j^{\left(  2\right)  }(p_{2})=\frac{ip}{2m}(1-e^{i\theta
}). $}$.$ The transfer 4-momentum, carried by the gauge or scalar mediators,
is: $k^{\mu}=P_{1}^{\mu}-P_{1}^{^{\prime}\mu}=(0,p(1-\cos\theta),-p\sin
\theta),$ whereas $\theta$ is the scattering angle (in the CM frame).

Considering the normalization condition satisfied by the spinors written in
eq. (\ref{spinor}), the scattering amplitude associated with the scalar sector
can be readily evaluated,
\begin{equation}
\mathcal{M}_{scalar}=y^{2}\frac{\left[  k^{2}-s^{2}\right]  }{\left[
k^{4}-\left(  s^{2}-v.v\right)  k^{2}-\left(  v.k\right)  ^{2}\right]  }.
\label{Mscalar1}%
\end{equation}
which in the case of a purely timelike background, $v^{\mu}=($v$_{0}$,
$\overrightarrow{0}),$ takes on the following form:%

\begin{equation}
\mathcal{M}_{scalar}=-y^{2}\frac{\left[  \mathbf{k}^{2}+s^{2}\right]
}{\mathbf{k}^{2}\left[  \mathbf{k}^{2}+w^{2}\right]  }, \label{Mscalar2}%
\end{equation}
where: $w^{2}=(s^{2}-$v$_{0}^{2}),$ and it was used the general expression for
the transfer momentum, $k^{\mu}=(0,\mathbf{k})$.

In connection with the gauge sector, only six terms of the gauge propagator
contribute to the scattering amplitude ($\theta^{\mu\nu},S^{\mu\nu}%
,\Lambda^{\mu\nu},T^{\mu}T^{\nu},Q^{\mu\nu},Q^{\nu\mu}),$ as a consequence of
the current-conservation law ($k_{\mu}J^{\mu}=0).$ The first two terms
provide, in the non-relativistic limit, the Maxwell-Chern-Simons
(MCS)\ scattering amplitude, already carried out in refs. \cite{MCS}:
\begin{equation}
\mathcal{M}_{MCS}=e^{2}\left\{  \left(  1-\frac{s}{m}\right)  \frac
{1}{\mathbf{k}^{2}+s^{2}}-\frac{2s}{m}\frac{i\overrightarrow{k}\times
\overrightarrow{p}}{\mathbf{k}^{2}(\mathbf{k}^{2}+s^{2})}\right\}  .
\label{MCS}%
\end{equation}
The total current-current amplitude mediated by the massive gauge particle
corresponds to the sum of four contributions,
\[
\mathcal{M}_{gauge}=\mathcal{M}_{MCS}+\text{ }\mathcal{M}_{\Lambda
}+\mathcal{M}_{TT}+\mathcal{M}_{QQ},
\]
where the terms $\mathcal{M}_{\Lambda},\mathcal{M}_{TT},\mathcal{M}_{QQ}$ lead
to\thinspace background-depending corrections to the MCS-amplitude. To
evaluate these three last terms, one first writes the following
current-current amplitudes:
\begin{align*}
j^{\mu}(p_{1})(T_{\mu}T_{\nu})j^{\nu}(p_{2})  &  =-2\frac{p^{4}}{m}%
\text{v}_{0}^{2}e^{i\theta}[1-\cos\theta+\sin^{2}\theta];\\
j^{\mu}(p_{1})\text{ }(\Lambda_{\mu v})j^{\nu}(p_{2})  &  =\text{v}_{0}^{2};\\
j^{\mu}(p_{1})\text{ }(Q_{\mu\nu}-Q_{\nu\mu})j^{\nu}(p_{2})  &  =2\frac{p^{2}%
}{m}\text{v}_{0}^{2}[1-\cos\theta-i\sin\theta];.
\end{align*}
The first term does not contribute to the interaction potential as long as one
works in the nonrelativistic approximation $(p^{2}\ll m^{2})$. The other two
terms lead to relevant contributions to the total amplitude scattering, namely:%

\begin{equation}
\text{ }\mathcal{M}_{\Lambda}=-\frac{e^{2}s^{2}\text{v}_{0}^{2}}%
{\mathbf{k}^{2}[\mathbf{k}^{2}+s^{2}][\mathbf{k}^{2}+w^{2}]},\mathcal{M}%
_{QQ}=\frac{e^{2}s\text{v}_{0}^{2}}{m}\frac{1}{[\mathbf{k}^{2}+s^{2}%
][\mathbf{k}^{2}+w^{2}]}\left\{  1-\frac{2i\overrightarrow{k}\times
\overrightarrow{p}}{\mathbf{k}^{2}}\right\}  ,
\end{equation}
where $\overrightarrow{p}$ $=\frac{1}{2}(\overrightarrow{p}_{1}%
-\overrightarrow{p}_{2})$ is defined in terms of the momenta $\overrightarrow
{p}_{1},\overrightarrow{p}_{2}$ of the incoming electrons.

\section{The electron-electron interaction potential}

\subsection{The scalar potential}

According to the Born approximation, the scalar interaction potential is given
by the Fourier transform of the scattering amplitude (\ref{Mscalar2}), that
is: $V_{scalar}(r)=\frac{1}{\left(  2\pi\right)  ^{2}}\int\mathcal{M}%
_{scalar}e^{i\overrightarrow{k}.\overrightarrow{r}}d^{2}\overrightarrow{k}$.
This integral can be exactly solved, resulting in the following expression:
\begin{equation}
V_{scalar}(r)=-\frac{y^{2}}{\left(  2\pi\right)  }\left\{  \left[
1+\frac{s^{2}}{w^{2}}\right]  K_{0}(sr)-\frac{s^{2}}{w^{2}}\ln r\right\}  .
\label{Vscalar}%
\end{equation}
This potential reveals to be attractive near the origin and repulsive whenever
the logarithmic term overcomes the Bessel-like one. Near the origin, the
potential exhibits a genuine logarithmic behavior, once $K_{0}(x)\rightarrow
-\ln x$ $($for $x\rightarrow0)$. Far from the origin, the bessel function
decays exponentially whereas the second term increases logarithmically. In
(1+2)-dimensions, the logarithmic behavior is an outcome consistent with an
unscreened interaction. Hence, the potential here obtained, at the level of a
dynamical configuration, confirms the annihilation of the screening derived in
ref. \cite{Manojr2}, at the level of a static evaluation. The result exhibited
in eq.(\ref{Vscalar}) reflects the pole structure of the scalar amplitude,
which possesses a massless $(1/\mathbf{k}^{2})$ and a massive pole $\left(
1/[\mathbf{k}^{2}+w^{2}]\right)  $. The existence of the massless pole is
ascribed to the fact the Chern-Simons pole $k^{2}=s^{2}$ to be deprived from
dynamics \cite{Manojr1}.

\subsection{The gauge potential}

Carrying out the Fourier transform on the $\mathcal{M}_{MCS}$-amplitude, the
corresponding Maxwell-Chern-Simons potential appears:
\begin{equation}
V_{MCS}(r)=\frac{e^{2}}{\left(  2\pi\right)  }\left[  \left(  1-\frac{s}%
{m}\right)  K_{0}(sr)-\frac{2}{ms}[1-srK_{1}(sr)]\frac{l}{r^{2}}\right]  ,
\label{VMCS}%
\end{equation}
where $l=\overrightarrow{r}\times\overrightarrow{p}$ is the angular momentum
(a scalar in a two-dimensional space).

The interaction potential associated with the amplitudes $\mathcal{M}%
_{\Lambda},\mathcal{M}_{QQ},$ can be also obtained from exact Fourier
transform, resulting in the following expressions:%

\begin{equation}
V_{\Lambda}(r)=\frac{e^{2}}{\left(  2\pi\right)  }\left\{  \frac{\text{v}%
_{0}^{2}}{w^{2}}\ln r+\frac{s^{2}}{w^{2}}K_{0}(wr)-K_{0}(sr)\right\}  ,
\end{equation}%

\begin{equation}
V_{QQ}(r)=\frac{e^{2}}{\left(  2\pi\right)  }\left\{  \frac{s}{m}\left[
K_{0}(wr)-K_{0}(sr)\right]  -\frac{2s}{m}\frac{l}{r^{2}}\left[  \frac
{\text{v}_{0}^{2}}{s^{2}w^{2}}-\frac{1}{w}rK_{1}(sr)+\frac{1}{s}%
rK_{1}(wr)\right]  \right\}  .
\end{equation}

The total gauge interaction potential, $V_{gauge}(r)=V_{MCS}+V_{\Lambda
}+V_{QQ},$ takes on the final form:
\begin{align}
V_{gauge}(r)  &  =\frac{e^{2}}{\left(  2\pi\right)  }\biggl\{-2(s/m)K_{0}%
(sr)+[s/m+s^{2}/w^{2}]K_{0}(wr)+\left(  \text{v}_{0}^{2}/w^{2}\right)  \ln
r\nonumber\\
&  -\frac{2}{ms}\frac{l}{r^{2}}\left[  (1+\text{v}_{0}^{2}/w^{2}%
)-(s^{2}/w)rK_{1}(sr)\right]  \biggr\}. \label{Vgauge}%
\end{align}

It is instructive to notice that one has $V_{\Lambda},V_{QQ}$ $\rightarrow0 $
in the limit of a vanishing background (v$_{0}\rightarrow0),$ recovering the
pure MCS result, given by eq. (\ref{VMCS}). Obviously, this is an expected
outcome, since both $V_{\Lambda},V_{QQ}$ are potential contributions induced
merely by the presence of the background. Taking the limit $r\rightarrow0$ on
the expression (\ref{Vgauge}), one then determines the potential behavior near
the origin, that is%

\begin{equation}
V_{gauge}(r)\simeq\frac{e^{2}}{\left(  2\pi\right)  }\biggl\{%
C-(1-s/m-2ls/m)\ln r\biggr\} \label{Vgauge2}%
\end{equation}
where $C$ is a constant$.$ Far from the origin, just the logarithmic term
remains as dominant, so that:
\begin{equation}
V_{gauge}(r)\simeq\frac{e^{2}}{\left(  2\pi\right)  }\left[  \frac
{\text{v}_{0}^{2}}{w^{2}}\right]  \ln r. \label{Vgauge3}%
\end{equation}
Eqs. (\ref{Vgauge2}), (\ref{Vgauge3}) show that the gauge potential behaves
logarithmically near and away from the origin, which puts again in evidence
the annihilation of the screening \cite{Manojr2}, now manifest at the level of
a dynamical evaluation. In the limit $r\rightarrow0,$ this potential may be
attractive (for $s>m/(1+l))$ or repulsive (for $s<m/(1+l))$. In this paper one
assumes $s^{2}>$ v$_{0}^{2}$, so that in the limit $r\rightarrow\infty$ the
potential behaves repulsively. In the case\ $s>m/(1+l)$, there exists a region
in which the potential is negative, a\ necessary premise for the formation of
electron-electron bound states. For the case $s<m/(1+l),$ in which the
potential is repulsive near and far the origin, just a graphical analysis can
efficiently reveal the existence of a well (negative) region.

The real interaction corresponds to\ the total interaction potential, which
comprises the gauge and the scalar contributions: $V(r)=V_{scalar}+V_{gauge}.
$ This total potential turns out attractive at the regions in which the
negative scalar potential overcomes the repulsive gauge contribution, and at
the regions for which the gauge potential is also negative. One then verifies
that the total potential can always be negative at some region, which is a
relevant result concerning the possibility of obtaining $e^{-}e^{-}$ bound
states in the framework of this particular model.

An important comparison to be made is allusive to the attractiveness of the
gauge potential. In the case of the pure MCS potential, given by eq.
(\ref{VMCS}),\ one must be careful in order to avoid a misleading
interpretation of its low-energy behavior \cite{MCS}. In such a regime, one
must consider not only the centrifugal barrier term $\left(  l^{2}%
/mr^{2}\right)  $, but also the gauge invariant $A^{2}-$term coming from the
Pauli equation,
\[
\left[  \frac{(p-eA)^{2}}{m_{e}}+e\phi(r)-\frac{\overrightarrow{\sigma
}.\overrightarrow{B}}{m_{e}}\right]  \Psi(r,\phi)=E\Psi(r,\phi),
\]
which rules the nonrelativistic behavior of a system in the presence of an
electromagnetic field. The Laplacian operator, $\left[  \frac{\partial^{2}%
}{\partial r^{2}}+\frac{1}{r}\frac{\partial}{\partial r}+\frac{1}{r^{2}}%
\frac{\partial^{2}}{\partial\phi^{2}}\right]  ,$ corresponding to the $p^{2}$
term, acts on the total wavefunction $\Psi(r,\phi)=R_{nl}(r)e^{i\phi l},$
generating \ the repulsive centrifugal barrier term, $l^{2}/\left(
mr^{2}\right)  .$ On the other hand, the $A^{2}$-term is essential to ensure
the gauge invariance of a gauge mode in the nonrelativistic domain. This term
does not appear\thinspace in the context of a nonperturbative low-energy
evaluation, for the same is associated with two-photon exchange processes (see
Hagen and Dobroliubov \cite{MCS}). However, such a term must be suitably added
up in order to assure the gauge invariance as well as to circumvent spurious
behaviour concerning the low-energy potential.

In the presence of these two terms, the MCS\ potential reveals to be really
repulsive instead of attractive. Hence, to correctly analyze the low-energy
behavior of the gauge potential, it is necessary to add up the centrifugal
barrier and the $A^{2}$ terms\footnote{The vector potential, $A(r),$ was not
determined in ref. \cite{Manojr2}, but it can be evaluated starting from the
following coupled equations: $\nabla^{2}(\nabla^{2}-s^{2})\overrightarrow{A}%
-$v$_{0}\nabla^{2}\nabla^{\ast}\varphi=s\nabla^{\ast}\rho,$ v$_{0}\nabla\times
A-\nabla^{2}\varphi=0,$ derived in the static limit. The solution of these
equations provides the required solution for the vector potential (in the case
of pure a time-like background): $\overrightarrow{A}(r)=\frac{e}{2\pi}%
\frac{s^{2}}{w^{2}}[1-wrK_{1}(wr)]\overset{\wedge}{r^{\ast}}.$ In ($1+2)$
dimensions, the dual of a 2-vector is defined as $\left(  E^{i}\right)
^{\ast}=\epsilon_{ij}E^{j}\longrightarrow\overrightarrow{E}^{\ast}%
=(E_{y},-E_{x}),$ where one adopts the following convection: $\epsilon
_{012}=\epsilon^{012}=\epsilon_{12}=\epsilon^{12}=1.$} to the gauge potential
previously obtained, leading to the following low-energy effective
potential:\
\[
V_{eff}(r)=V_{gauge}(r)+\frac{l^{2}}{mr^{2}}+\left(  \frac{e}{2\pi}\right)
^{2}\left(  \frac{s^{2}}{w^{2}}\right)  ^{2}[1-wrK_{1}(wr)]^{2}.
\]
The possibility for formation of electron-electron bound states is associated
with the existence of a region in which the effective potential is negative.
The figure below shows that this requirement is perfectly fulfilled for some
parameters values:%

%TCIMACRO{\FRAME{ftbpFU}{4.0612in}{3.0588in}{0pt}{\Qcb{ Effective potential for
%the following parameter values: $s=10,m=2000,$v$_{0}=8,l=1.$}}{}%
%{moller.eps}{\special{ language "Scientific Word";  type "GRAPHIC";
%width 4.0612in;  height 3.0588in;  depth 0pt;  original-width 3.9851in;
%original-height 2.9948in;  cropleft "0";  croptop "1";  cropright "1";
%cropbottom "0";  filename 'moller.eps';file-properties "XNPEU";}}}%
%BeginExpansion
\begin{figure}
[ptb]
\begin{center}
\includegraphics[
height=3.0588in,
width=4.0612in
]%
{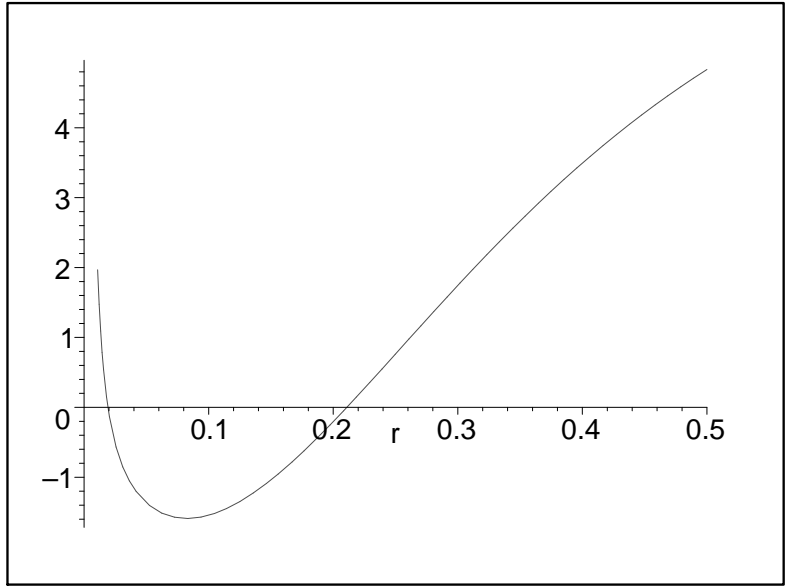}%
\caption{ Effective potential for the following parameter values:
$s=10,m=2000,$v$_{0}=8,l=1.$}%
\end{center}
\end{figure}
%EndExpansion

\section{Concluding Remarks}

In this work, one has considered the M\"{o}ller scattering in the context of a
planar Lorentz-violating Maxwell-Chern-Simons electrodynamics defined in a
pure timelike background. The interaction potential was calculated as the
Fourier transform of the scattering amplitude (Born approximation) carried out
in the non-relativistic limit. The interaction potential exhibits two distinct
contributions: the attractive scalar one (stemming form the Yukawa exchange)
and the gauge one (mediated by the MCS-Proca gauge field). The scalar Yukawa
interaction, as expected, turns out to be always negative. This makes feasible
a global attractive potential, regardless the character (repulsive or
attractive) of the gauge potential. In practice, such an interaction may be
identified with phonon exchange processes, which represent physical
excitations in several systems of interest. As for the gauge interaction, it
is composed by a pure MCS potential corrected by background-depending
contributions, which impose relevant physical modifications. The absence of
screening, first observed in ref. \cite{Manojr2}, becomes now manifest in the
context of a dynamical computation (by means of an ubiquitous logarithmic
term), confirming the conclusion that $k^{2}=s^{2}$ is not a dynamical pole
\cite{Manojr1}. The background-depending corrections are such that they lead
to an attractive gauge potential for some values of the parameters, which
constitutes a promising result in connection with the possibility of obtaining
the formation of Cooper pairs. This possibility can be appropriately checked
up by means of a quantum-mechanical numerical analysis of the interaction
potential here derived, which should be performed by means of the numerical
solution of the Schr\"{o}dinger equation. Such analysis must provide the
corresponding $e^{-}e^{-}$ binding energy once one takes suitable values for
the parameters (in accordance with the scale of low-energy excitations typical
in condensed matter systems).

A natural extension of this work consists in studying the interaction
potential for the case of a purely spacelike background\cite{Manojr4}. Such an
evaluation will certainly reveal an anisotropic potential in relation to the
privileged direction fixed by the background, which may lead to an attractive
interaction as well as an anisotropic $e^{-}e^{-}$ order parameter.

In (1+2) dimensions, the purely Coulombian interaction is associated with a
logarithmic dependence, which implies in a confining rather than a binding
behavior, which is ubiquitous in the results of this paper. However, the same
can be eliminated if the gauge field exhibits an additional mass component, as
the Proca term. Indeed, in a recent work \cite{Manojr3} it was accomplished
the dimensional reduction of an Abelian-Higgs Lorentz-violating model endowed
with the CFJ term, resulting in a planar Maxwell-Chern-Simons-Proca
electrodynamics coupled to a massive Klein-Gordon field $\left(
\varphi\right)  $. A particular feature of this kind of Higgs model is the
presence of totally screened modes: all its physical excitations are massive,
which yields screened interactions.\textbf{\ }The consideration of the
M\"{o}ller scattering in this framework will lead to an entirely shielded
interaction potential, once the logarithmic term should be suitably replaced
by a K$_{0}$ function.

Acknowledgement: \ The author is grateful to J. A. Helay\"{e}l-Neto for
reading and discussing this manuscript.

\end{document}